%Paper: hep-th/9205091
%From: RRK@phys.tamu.edu
%Date: Tue, 26 May 1992 16:43:49 CDT

\input harvmac.tex
\Title{CTP/TAMU-38/92}{{Monopoles and Instantons in String Theory}
\footnote{$^\dagger$}{Work supported in part by NSF grant PHY-9106593.}}
%For more complicated situations, substitute for {\it either\/} argument:
%\Title{\vbox{\baselineskip12pt\hbox{HUTP-88/A000}\hbox{SLAC-PUB 88-001}
%		\hbox{photocopy at own risk}}}
%{\vbox{\centerline{This title is too long to fit}
%	\vskip2pt\centerline{comfortably on one line*}}}
%   \footnote{}{*optional footnote on title}

\centerline{
Ramzi~R.~ Khuri\footnote{$^*$}{Supported by a World Laboratory Fellowship.}}
\bigskip\centerline{Center for Theoretical Physics}
\centerline{Texas A\&M University}\centerline{College Station, TX 77843}

\vskip .3in
In recent work, several classes of solitonic solutions of string theory
with higher-membrane structure have been obtained. These solutions can be
classified according to the symmetry possessed by the solitons in the subspace
of the spacetime transverse to the membrane. Solitons with four-dimensional
spherical symmetry represent instanton solutions in string theory,
while those with three-dimensional spherical symmetry represent magnetic
monopole-type solutions. For both of these classes, we discuss bosonic as well
as heterotic solutions.

\Date{5/92}
%\draft
\def\wbg{\overline{\beta}^G_{\mu\nu}}
\def\wbb{\overline{\beta}^B_{\mu\nu}}
\def\wbd{\overline{\beta}^\Phi}
\def\rbg{\beta^G_{\mu\nu}}
\def\rbb{\beta^B_{\mu\nu}}
\def\rbd{\beta^\Phi}
\def\sqr#1#2{{\vbox{\hrule height.#2pt\hbox{\vrule width
.#2pt height#1pt \kern#1pt\vrule width.#2pt}\hrule height.#2pt}}}
\def\Box{\mathchoice\sqr64\sqr64\sqr{4.2}3\sqr33}
\def\rijkl{R^i{}_{jkl}}
\def\grijkl{\hat R^i{}_{jkl}}
\def\hstar {H^\ast_\mu}
\def\met {g_{\mu\nu}}

\lref\reyone {S.--J.~Rey {\it Axionic String Instantons
and their Low-Energy Implications}, Proceedings, Tuscaloosa 1989,
Superstrings and particle theory, p.291.}

\lref\reytwo {S.--J.~Rey, Phys. Rev. {\bf D43} (1991) 526.}

\lref\abenone{I.~Antoniadis, C.~Bachas, J.~Ellis and D.~V.~Nanopoulos,
Phys. Lett. {\bf B211} (1988) 393.}

\lref\abentwo{I.~Antoniadis, C.~Bachas, J.~Ellis and D.~V.~Nanopoulos,
Nucl. Phys. {\bf B328} (1989) 117.}

\lref\mtone{R.~R.~Metsaev and A.~A.~Tseytlin, Phys. Lett.
{\bf B191} (1987) 354.}

\lref\mttwo{R.~R.~Metsaev and A.~A.~Tseytlin,
Nucl. Phys. {\bf B293} (1987) 385.}

\lref\cfmp{C.~G.~Callan, D.~Friedan, E.~J.~Martinec
and M.~J.~Perry, Nucl. Phys. {\bf B262} (1985) 593.}

\lref\ckp{C.~G.~Callan,
I.~R.~Klebanov and M.~J.~Perry, Nucl. Phys. {\bf B278} (1986) 78.}

\lref\love{C.~Lovelace, Phys. Lett. {\bf B135} (1984) 75.}

\lref\fridven{B.~E.~Fridling and A.~E.~M.~Van de Ven,
Nucl. Phys. {\bf B268} (1986) 719.}

\lref\gepwit{D.~Gepner and E.~Witten, Nucl. Phys. {\bf B278} (1986) 493.}

\lref\quartet{D.~J.~Gross,
J.~A.~Harvey, E.~J.~Martinec and R.~Rohm, Nucl. Phys. {\bf B267} (1986) 75.}

\lref\dine{M.~Dine, Lectures delivered at
TASI 1988, Brown University (1988) 653.}

\lref\brone{E.~A.~Bergshoeff and M.~de Roo, Nucl.
Phys. {\bf B328} (1989) 439.}

\lref\brtwo{E.~A.~Bergshoeff and M.~de Roo, Phys. Lett. {\bf B218} (1989) 210.}

\lref\chsone{C.~G.~Callan, J.~A.~Harvey and A.~Strominger, Nucl. Phys.
{\bf B359} (1991) 611.}

\lref\chstwo{C.~G.~Callan, J.~A.~Harvey and A.~Strominger, Nucl. Phys.
{\bf B367} (1991) 60.}

\lref\bpst{A.~A.~Belavin, A.~M.~Polyakov, A.~S.~Schwartz and Yu.~S.~Tyupkin,
Phys. Lett. {\bf B59} (1975) 85.}

\lref\thooft{G.~'t~Hooft, Nucl. Phys. {\bf B79} (1974) 276.}

\lref\hoofan{G.~'t~Hooft, Phys. Rev. Lett., {\bf 37} (1976) 8.}

\lref\wil{F.~Wilczek, in {\it Quark confinement and field theory},
Ed. D.~Stump and D.~Weingarten, John Wiley and Sons, New York (1977).}

\lref\cofa{E.~Corrigan and D.~B.~Fairlie, Phys. Lett. {\bf B67} (1977) 69.}

\lref\jackone{R.~Jackiw, C.~Nohl and C.~Rebbi, Phys. Rev. {\bf D15} (1977)
1642.}

\lref\jacktwo{R.~Jackiw, C.~Nohl and C.~Rebbi, in {\it Particles and
Fields}, Ed. David Boal and A.~N.~Kamal, Plenum Publishing Co., New York
(1978), p.199.}

\lref\rkinst{R.~R.~Khuri, Phys. Lett. {\bf B259} (1991) 261.}

\lref\rkscat{C.~G.~Callan and R.~R.~Khuri, Phys. Lett. {\bf B261} (1991) 363.}

\lref\rkmant{R.~R.~Khuri, {\it Manton Scattering of String Solitons}
PUPT-1270 (to appear in Nucl. Phys. {\bf B}).}

\lref\rkdg{R.~R.~Khuri, {\it Some Instanton Solutions in String
Theory} to appear in Proceedings of the XXth International Conference on
Differential Geometric Methods in Theoretical Physics, World Scientific,
October 1991.}

\lref\rkthes{R.~R.~Khuri, {\it Solitons and Instantons in String Theory},
 Princeton University Doctoral Thesis, August 1991.}

\lref\rksing{M.~J.~Duff, R.~R.~Khuri and J.~X.~Lu, {\it String and
Fivebrane Solitons: Singular or Non-singular?}, Texas A\&M preprint,
CTP/TAMU-89/91 (to appear in Nucl. Phys. {\bf B}).}

\lref\rkorb{R.~R.~Khuri and H.~S.~La, {\it Orbits of a String around a
Fivebrane}, Texas A\&M preprint, CTP/TAMU-95/91
(submitted to Phys. Rev. Lett.).}

\lref\rkmot{R.~R.~Khuri and H.~S.~La, {\it String Motion in Fivebrane
Geometry}, Texas A\&M preprint, CTP/TAMU-98/91 (submitted to Nucl. Phys. B).}

\lref\rkmonex{R.~R.~Khuri {\it A Heterotic Multimonopole Solution},
Texas A\&M preprint, CTP/TAMU-35/92.}

\lref\rkmono{R.~R.~Khuri {\it A Multimonopole Solution in
String Theory}, Texas A\&M preprint, CTP/TAMU-33/92.}

\lref\rkmscat{R.~R.~Khuri {\it Scattering of String Monopoles},
Texas A\&M preprint, CTP/TAMU-34/92.}

\lref\ginsp{P.~Ginsparg, Lectures delivered at
Les Houches summer session, June 28--August 5, 1988.}

\lref\swzw {W.~Boucher, D.~Friedan and A.~Kent, Phys. Lett.
{\bf B172} (1986) 316.}

\lref\dghrr{A.~Dabholkar, G.~Gibbons, J.~A.~Harvey and F.~Ruiz Ruiz,
Nucl. Phys. {\bf B340} (1990) 33.}

\lref\dabhar{A.~Dabholkar and J.~A.~Harvey,
Phys. Rev. Lett. {\bf 63} (1989) 478.}

\lref\prso{M.~K.~Prasad and C.~M.~Sommerfield, Phys. Rev. Lett. {\bf 35}
(1975) 760.}

\lref\jim{J.~A.~Harvey and J.~Liu, Phys. Lett. {\bf B268} (1991) 40.}

\lref\mantone{N.~S.~Manton, Nucl. Phys. {\bf B126} (1977) 525.}

\lref\manttwo{N.~S.~Manton, Phys. Lett. {\bf B110} (1982) 54.}

\lref\mantthree{N.~S.~Manton, Phys. Lett. {\bf B154} (1985) 397.}

\lref\atiyah{M.~F.~Atiyah and N.~J.~Hitchin, Phys. Lett. {\bf A107}
(1985) 21.}

\lref\atiyahbook{M.~F.~Atiyah and N.~J.~Hitchin, {\it The Geometry and
Dynamics of Magnetic Monopoles}, Princeton University Press, 1988.}

\lref\strom{A.~Strominger, Nucl. Phys. {\bf B343} (1990) 167.}

\lref\gsw{M.~B.~Green, J.~H.~Schwartz and E.~Witten,
{\it Superstring Theory} vol. 1, Cambridge University Press (1987).}

\lref\polch{J.~Polchinski, Phys. Lett. {\bf B209} (1988) 252.}

\lref\dfluone{M.~J.~Duff and J.~X.~Lu, Nucl. Phys. {\bf B354} (1991) 141.}

\lref\dflutwo{M.~J.~Duff and J.~X.~Lu, Nucl. Phys. {\bf B354} (1991) 129.}

\lref\dfluthree{M.~J.~Duff and J.~X.~Lu, Phys. Rev. Lett. {\bf 66}
(1991) 1402.}

\lref\dflufour{M.~J.~Duff and J.~X.~Lu, Nucl. Phys. {\bf B357} (1991)
534.}

\lref\dfstel{M.~J.~Duff and K.~S.~Stelle, Phys. Lett. {\bf B253} (1991)
113.}

\lref\ferone{R.~C.~Ferrell and D.~M.~Eardley, Phys. Rev. Lett. {\bf 59}
(1987) 1617.}

\lref\fertwo{R.~C.~Ferrell and D.~M.~Eardley, {\it Slowly Moving
Maximally Charged Black Holes} in Frontiers in Numerical Relativity,
Cambridge University Press, 1987.}

\lref\gh{G.~W.~Gibbons and S.~W.~Hawking, Phys. Rev. {\bf D15}
(1977) 2752.}

\lref\ghp{G.~W.~Gibbons, S.~W.~Hawking and M.~J.~Perry, Nucl. Phys. {\bf B318}
(1978) 141.}

\lref\briho{D.~Brill and G.~T.~Horowitz, Phys. Lett. {\bf B262} (1991)
437.}

\lref\gidone{S.~B.~Giddings and A.~Strominger, Nucl. Phys. {\bf B306}
(1988) 890.}

\lref\gidtwo{S.~B.~Giddings and A.~Strominger, Phys. Lett. {\bf B230}
(1989) 46.}

\lref\raj{R.~Rajaraman, {\it Solitons and Instantons}, North Holland,
1982.}

\lref\chsw{P.~Candelas, G.~T.~Horowitz, A.~Strominger and E.~Witten,
Nucl. Phys. {\bf B258} (1984) 46.}

\lref\bogo{E.~B.~Bogomolnyi, Sov. J. Nucl. Phys. {\bf 24} (1976) 449.}

\lref\cogo{E.~Corrigan and P.~Goddard, Comm. Math. Phys. {\bf 80} (1981)
575.}

\lref\wardone{R.~S.~Ward, Comm. Math. Phys. {\bf 79} (1981) 317.}

\lref\wardtwo{R.~S.~Ward, Comm. Math. Phys. {\bf 80} (1981) 563.}

\lref\wardthree{R.~S.~Ward, Phys. Lett. {\bf B158} (1985) 424.}

\lref\groper{D.~J.~Gross and M.~J.~Perry, Nucl. Phys. {\bf B226} (1983)
29.}

\lref\ash{{\it New Perspectives in Canonical Gravity}, ed. A.~Ashtekar,
Bibliopolis, 1988.}

\lref\lich{A.~Lichnerowicz, {\it Th\' eories Relativistes de la
Gravitation et de l'Electro-magnetisme}, (Masson, Paris 1955).}

\lref\goldstein{H.~Goldstein, {\it Classical Mechanics}, Addison-Wesley,
1981.}

\lref\dflufive{M.~J.~Duff and J.~X.~Lu, Class. Quant. Grav. {\bf 9}
(1992) 1.}

\lref\dflusix{M.~J.~Duff and J.~X.~Lu, Phys. Lett. {\bf B273} (1991)
409.}

\lref\hlp{J.~Hughes, J.~Liu and J.~Polchinski, Phys. Lett. {\bf B180}
(1986).}

\lref\town{P.~K.~Townsend, Phys. Lett. {\bf B202} (1988) 53.}

\lref\duff{M.~J.~Duff, Class. Quant. Grav. {\bf 5} (1988).}

\lref\rossi{P.~Rossi, Physics Reports, 86(6) 317-362.}

\lref\gksone{B.~Grossman, T.~W.~Kephart and J.~D.~Stasheff, Commun. Math.
Phys. {\bf 96} (1984) 431.}

\lref\gkstwo{B.~Grossman, T.~W.~Kephart and J.~D.~Stasheff, Commun. Math.
Phys. {\bf 100} (1985) 311.}

\lref\gksthree{B.~Grossman, T.~W.~Kephart and J.~D.~Stasheff, Phys. Lett.
{\bf B220} (1989) 431.}

\newsec{Introduction}

In recent work classical solitonic solutions of string theory with
higher-membrane structure have been investigated. These solutions
can be classified according to the symmetry the solitons possess in the
subspace
of spacetime transverse to the membrane. In this paper, we discuss two classes
of solutions, those with four-dimensional spherical symmetry, which possess
instanton structure, and those with three-dimensional spherical
symmetry, which represent magnetic monopole-type solutions in string theory.

For both instantons and monopoles, we review solutions in Yang-Mills field
theory as well as axionic solitonic solutions for the massless fields of the
bosonic string. In each case we combine the gauge theory solution with the
corresponding bosonic solution to obtain an exact multi-soliton
solution of heterotic string theory\quartet.

We begin section 2 with a review of the 't Hooft ansatz for the Yang-Mills
instanton\refs{\hoofan\wil\cofa\jackone{--}\jacktwo}. We then turn to the
axionic instanton solution first mentioned in \reyone.
This tree-level solution is extended in \rkinst\ to an exact
solution of bosonic string theory for the special case of a linear dilaton
wormhole solution\refs{\abenone,\abentwo}. Exactness is shown by combining
the metric and antisymmetric tensor in a generalized curvature, which
is written covariantly in terms of the tree-level dilaton field, and rescaling
the dilaton order by order in the parameter $\alpha'$. The corresponding
conformal field theory is written down.

An exact heterotic multi-soliton solution with YM instanton structure in
the four dimensional transverse space can be obtained\refs{\chsone,\chstwo} by
equating the curvature of the Yang-Mills gauge field with the generalized
curvature derived in \rkinst. This solution represents an exact extension of
the tree-level fivebrane solutions of \refs{\strom,\dfluone,\dflutwo}\ and
combines the gauge and axionic instanton structures.

In section 3 we turn to the three-dimensional (monopole) solutions. We first
discuss a multimonopole solution in YM field theory, which
arises from a modification of the 't Hooft ansatz for the four-dimensional
instanton\refs{\rkmono,\rkmonex}. We then mention the bosonic three-dimensional
solution obtained in \rkthes. We complete this section with a review of the
recently constructed exact multimonopole solution of heterotic string
theory\refs{\rkmono,\rkmonex}, which now combines the gauge and axionic
monopole structures. Unlike the heterotic instanton solution, this
solution does not lend itself easily to a CFT description. An interesting
aspect
of this string monopole solution, however, is that the divergences stemming
from the YM sector are precisely cancelled by those coming from the gravity
sector, thus resulting in a finite action solution.

We conclude in section 4 with a summary of these results and a brief
discussion.

\newsec{Four-Dimensional Instanton Solutions}

In this section, we discuss four-dimensional, or instanton solutions in
bosonic and heterotic string theory. We first summarize the 't Hooft ansatz
for the Yang-Mills instanton, and then write down the tree-level bosonic
axionic instanton solution of \reyone. An exact extension of this solution
can be obtained for the special case of a wormhole solution, and the
corresponding conformal field theory is written down\rkinst.
Finally, an exact multi-instanton solution of heterotic string theory is
obtained, combining the Yang-Mills gauge solution with the bosonic axionic
instanton\refs{\rkdg,\chsone,\chstwo}.

Consider the four-dimensional Euclidean action
\eqn\ymact{S=-{1\over 2g^2}\int d^4x {\rm Tr} G_{\mu\nu}G^{\mu\nu},
\qquad\qquad \mu,\nu =1,2,3,4.}
For gauge group $SU(2)$, the fields may be written as $A_\mu=(g/2i)
\sigma^a A_\mu^a$ and $G_{\mu\nu}=(g/2i)\sigma^a G_{\mu\nu}^a$\ \
(where $\sigma^a$, $a=1,2,3$ are the $2\times 2$ Pauli matrices).
The equation of motion derived from this action is solved by the
't Hooft ansatz\refs{\hoofan\wil\cofa\jackone{--}\jacktwo}
\eqn\hfanstz{A_\mu=i \overline{\Sigma}_{\mu\nu}\partial_\nu \ln f,}
where $\overline{\Sigma}_{\mu\nu}=\overline{\eta}^{i\mu\nu}(\sigma^i/2)$
for $i=1,2,3$, where
\eqn\hfeta{\eqalign{\overline{\eta}^{i\mu\nu}=-\overline{\eta}^{i\nu\mu}
&=\epsilon^{i\mu\nu},\qquad\qquad \mu,\nu=1,2,3,\cr
&=-\delta^{i\mu},\qquad\qquad \nu=4 \cr}}
and where $f^{-1}\Box\ f=0$. The ansatz for the anti-self-dual solution
is similar, with the $\delta$-term in \hfeta\ changing sign.
To obtain a multi-instanton solution, one solves for $f$ in the
four-dimensional space to obtain
\eqn\finst{f=1+\sum_{i=1}^N{\rho_i^2\over |\vec x - \vec a_i|^2},}
where $\rho_i^2$ is the instanton scale size and $\vec a_i$ the location in
four-space of the $i$th instanton. We will return to the 't Hooft ansatz
when we consider a superstring model with YM coupling
(the heterotic string\quartet).

We now turn to the bosonic axionic instanton solution
considered in \rkinst. We first derive the tree-level solution of \reyone\
and then extend the single instanton wormhole solution to $O(\alpha')$ in
the massless fields. For this purpose we use the theorem of equivalence of the
massless string field equations to the sigma-model Weyl invariance
conditions (demonstrated to two-loop order by Metsaev and
Tseytlin\refs{\mtone,\mttwo}), which require the Weyl
anomaly coefficients $\wbg$, $\wbb$ and $\wbd$ to vanish identically to
the appropriate order in the parameter $\alpha'$.
The two-loop solution obtained by this method suggests a
representation of the sigma model as the product of a WZW\gepwit\ model
and a one-dimensional CFT (a Feigin-Fuchs Coulomb gas)\reyone.
This representation allows us to obtain an exact solution.

The bosonic sigma model action can be written as\love
\eqn\sigmod{I={1\over 4\pi\alpha'}\int d^2x\left(\sqrt{\gamma}\gamma^{ab}
\partial_ax^\mu\partial_bx^\nu\met+i\epsilon^{ab}\partial_ax^\mu\partial_b
x^\nu B_{\mu\nu}+\alpha'\sqrt{\gamma}R^{(2)}\phi\right),}
where $\met$ is the sigma model metric, $\phi$ the dilaton and $B_{\mu\nu}$
the antisymmetric tensor, and where $\gamma_{ab}$ is the worldsheet metric
and $R^{(2)}$ the two-dimensional curvature.
The Weyl anomaly coefficients are given by\refs{\mtone,\mttwo}
\eqn\weyl{\eqalign{\wbg&=\rbg+2\alpha'\nabla_\mu\nabla_\nu\phi+\nabla_{(\mu}
W_{\nu )},\cr\wbb&=\rbb+\alpha' {H_{\mu\nu}}^\lambda\partial_\lambda\phi
+{1\over 2}{H_{\mu\nu}}^\lambda W_\lambda,\cr
\wbd&=\rbd+\alpha'(\partial\phi)^2+{1\over 2}W^\lambda\partial_\lambda\phi,
,\cr}}
where $\rbg$, $\rbb$ and $\rbd$ are the RG $\beta$ functions and where
$H_{\mu\nu\lambda}=\partial_{[\mu}B_{\nu\lambda]}$ and
$W_\mu=-(\alpha'^2/24)\nabla_\mu H^2$.

We first show that for any dilaton function satisfying
$e^{-2\phi}\Box\ e^{2\phi}=0$ with
\eqn\sansatz{\eqalign{\met&=e^{2\phi}\delta_{\mu\nu}\qquad \mu,\nu=1,2,3,4,\cr
g_{ab}&=\delta_{ab}\qquad\quad   a,b=5,...,26,\cr
H_{\mu\nu\lambda}&=\pm\epsilon_{\mu\nu\lambda\sigma}\partial^\sigma\phi
\qquad \mu,\nu,\lambda,\sigma=1,2,3,4\cr}}
the $O(\alpha')$ Weyl anomaly coefficients vanish identically.

We define a generalized curvature $\grijkl$ in terms of the standard
curvature $\rijkl$ and $H_{\mu\alpha\beta}$\fridven:
\eqn\gcurv{\grijkl=\rijkl+{1\over
2}\left(\nabla_lH^i{}_{jk}-\nabla_kH^i{}_{jl}\right)
+{1\over 4}\left(H^m{}_{jk}H^i{}_{lm}-H^m{}_{jl}H^i{}_{km}\right).}
One can also define $\grijkl$ as the Riemann tensor generated
by the generalized Christoffel symbols $\hat\Gamma^\mu_{\alpha\beta}$,
where  $\hat\Gamma^\mu_{\alpha\beta}=\Gamma^\mu_{\alpha\beta}
-(1/2) H^\mu{}_{\alpha\beta}$.

We follow Metsaev and Tseytlin's computation of the renormalization
group beta functions for the general sigma-model and combine dimensional
regularization and the minimal subtraction scheme with the following
generalized prescription for contraction of $\epsilon^{ab}$ tensors\mtone:
\eqn\reg{\epsilon^{ab}\epsilon^{cd}=f(d)\left(\delta^{ac}\delta^{bd}
-\delta^{ad}\delta^{bc}\right),}
where $f(d)=1-f_1\epsilon+O(\epsilon^2)$ and $\epsilon=d-2$.
We note that the precise form of the renormalization group beta functions at
two-loop order is not scheme-independent but depends on the choice of
$f_1$. Here we set $f_1=-1$, for which Metsaev and Tseytlin obtain the
following two-loop expressions for the Weyl anomaly
coefficients\refs{\mtone,\mttwo}:
\eqn\weycos{\eqalign{\wbg&=\alpha'(\hat R_{(\mu\nu)}+2\nabla_\mu
\nabla_\nu\phi)\cr&+{\alpha'^2\over
2}\left(\hat R^{\alpha\beta\gamma}{}_{(\mu}\hat R_{\nu)\alpha\beta\gamma}
-{1\over 2}\hat R^{\beta\gamma\alpha}{}_{(\mu}\hat
R_{\nu)\alpha\beta\gamma}+{1\over 2}\hat R_{\alpha(\mu\nu)\beta}
(H^2)^{\alpha\beta}\right)+\nabla_{(\mu}W_{\nu)},\cr
\wbb&=\alpha'(\hat R_{[\mu\nu]}+H_{\mu\nu}{}^\lambda\partial_\lambda\phi)
\cr&+{\alpha'^2\over
2}\left(\hat R^{\alpha\beta\gamma}{}_{[\mu}\hat R_{\nu]\alpha\beta\gamma}
-{1\over 2}\hat R^{\beta\gamma\alpha}{}_{[\mu}\hat
R_{\nu]\alpha\beta\gamma}+{1\over 2}\hat R_{\alpha[\mu\nu]\beta}
(H^2)^{\alpha\beta}\right)+{1\over 2}H_{\mu\nu}{}^\lambda W_\lambda
,\cr
\wbd&={D\over 6}-{\alpha'\over 2}\left(\nabla^2\phi-2(\partial\phi)^2+{1\over
12}H^2\right)\cr&+{\alpha'^2\over 16}\left(2(H^2)^{\mu\nu}\nabla_\mu
\nabla_\nu\phi+R^2_{\lambda\mu\nu\rho}-{11\over 2}RHH
+{5\over 24}H^4+{11\over 8}(H^2_{\mu\nu})^2+{4\over 3}\nabla H\cdot
\nabla H\right)\cr&+{1\over 2}W^\lambda\partial_\lambda\phi,\cr}}
where $\nabla H\cdot\nabla H\equiv\nabla_\alpha H_{\beta\gamma\delta}
\nabla^\alpha H^{\beta\gamma\delta}.$
Unless otherwise indicated, all expressions are written to two loop
order in the beta-functions, which corresponds to $O(\alpha')$ in the action.
Also, all indices are in the curved four-space, as it is clear that the
flat dimensions do not contribute.

The crucial observation for obtaining higher-loop and even exact solutions
is the following. For any solution of the form \sansatz,
we can express the generalized curvature in covariant form in terms of
the dilaton field $\phi$:
\eqn\gcurvphi{\grijkl=\delta_{il}\nabla_k\nabla_j\phi
-\delta_{ik}\nabla_l\nabla_j\phi+\delta_{jk}\nabla_l\nabla_i\phi
-\delta_{jl}\nabla_k\nabla_i\phi\pm\epsilon_{ijkm}\nabla_l\nabla_m\phi
\mp\epsilon_{ijlm}\nabla_k\nabla_m\phi,}
It follows from \gcurvphi\ that
\eqn\gricci{\eqalign{\hat R_{(\mu\nu)}&=-2\nabla_\mu\nabla_\nu\phi,\cr
\hat R_{[\mu\nu]}&=0.\cr}}
It also follows from \sansatz\ that
\eqn\zeroterms{\eqalign{\nabla^2\phi&=0,\cr{H_{\mu\nu}}^\lambda
\partial_\lambda\phi&=0,\cr H^2&=24(\partial\phi)^2.\cr}}
{}From \gricci\ and \zeroterms\ it follows that the $O(\alpha')$ terms
in the Weyl anomaly coefficients in \weycos\ vanish identically for the
ansatz \sansatz. A tree-level multi-instanton solution is therefore
given by \sansatz\ with the dilaton given by
\eqn\multy{e^{2\phi}=C+\sum_{i=1}^N {Q_i\over |\vec x -\vec a_i|^2},}
where $Q_i$ is the charge and $\vec a_i$ the location in the four-space
$(1234)$ of the $i$th instanton. We call $(1234)$ the transverse space,
as the solitons have the structure of $21+1$-dimensional objects embedded
in a $26$-dimensional spacetime.

We now specialize to the spherically symmetric case of $e^{2\phi}={Q/r^2}$ in
\sansatz\ and determine the $O(\alpha')$ corrections to the massless fields in
\sansatz\ so that the Weyl anomaly coefficients vanish to $O(\alpha'^2)$.
For this solution we notice
\eqn\ddphi{\nabla_\mu\nabla_\nu\phi=0,}
and therefore from \gcurvphi
\eqn\gcurvzero{\grijkl=0,}
and we have what is called a ``parallelizable" space\refs{\mtone,\mttwo}.
To maintain a parallelizable space to $O(\alpha')$ we keep $\met$
and $H_{\alpha\beta\gamma}$ in their lowest order form and assume
that any corrections to \sansatz\ appear in the dilaton:
\eqn\tlgansatz{\eqalign{\phi&=\phi_0+\alpha'\phi_1+...\cr
e^{2\phi_0}&={Q\over r^2},\cr
\met&=e^{2\phi_0}\delta_{\mu\nu},\cr
H_{\mu\nu\lambda}&=\pm\epsilon_{\mu\nu\lambda\sigma}\partial^\sigma\phi_0
.\cr}}
It follows from \tlgansatz\ that $H^2=24(\partial\phi_0)^2=24/Q$ and
thus $W_\mu=0$. It follows from \gcurvzero\ that $\wbg$ and $\wbb$
vanish identically to two loop order and that
\eqn\dweylone{\eqalign{\wbd={D\over 6}+&\alpha'\left((\partial\phi)^2
-{1\over Q} \right)\cr
&+{\alpha'^2\over 16}\left(R^2_{\lambda\mu\nu\rho}-{11\over 2}RHH
+{5\over 24}H^4+{11\over 8}(H^2_{\mu\nu})^2+{4\over 3}\nabla H\cdot\nabla H
\right).\cr}}
We use the relations in equation (34) in \mtone\ for parallelizable spaces
and the observation that $(H^2_{\mu\nu})^2=2H^4=192/Q^2$ for our solution to
get the identities
\eqn\parallel{\eqalign{R^2_{\lambda\mu\nu\rho}&={1\over 8}H^4,\cr
RHH&={1\over 2}H^4,\cr\nabla H\cdot\nabla H&=0 .\cr}}
\dweylone\ then simplifies further to
\eqn\dweyltwo{\wbd={D\over 6}+\alpha'\left((\partial\phi)^2-{1\over Q}\right)
+2{\alpha'^2\over Q^2}.}
The lowest order term in $\wbd$ is proportional to the central charge
and the $O(\alpha')$ terms vanish identically. With the choice
$\overrightarrow\nabla\phi_1=-(1/Q)\overrightarrow\nabla\phi_0$, the
$O(\alpha'^2)$ terms also vanish identically.
The two-loop solution is then given by
\eqn\tlsansatz{\eqalign{e^{2\phi}&={Q\over r^{2(1-{\alpha'\over Q})}},\cr
\met&={Q\over r^2}\delta_{\mu\nu},\cr
H_{\mu\nu\lambda}&=\pm\epsilon_{\mu\nu\lambda\sigma}\partial^\sigma\phi_0,\cr}}
which corresponds to a simple rescaling of the dilaton.
A quick check shows that this solution has finite action near the
singularity.

We now rewrite $\wbd$ in \dweyltwo\ in the following suggestive form:
\eqn\dweylsplit{\eqalign{6\wbd&=\left(1+6\alpha'(\partial\phi)^2\right)
+\left(3-6{\alpha'\over Q}+12({\alpha'\over Q})^2\right)\cr&=4.\cr}}
The above splitting of the central charge $c=6\wbd$ suggests
the decomposition of the corresponding sigma model into the
product of a one-dimensional CFT (a Feigin-Fuchs Coulomb gas)
and a three-dimensional WZW model with an
$SU(2)$ group manifold \refs{\reyone,\mtone,\mttwo}.
This can be seen as follows.
Setting $u=\ln r$, we can rewrite \sigmod\ for our solution\reyone\
in the form $I=I_1+I_3$, where
\eqn\onecft{I_1={1\over 4\pi\alpha'}\int d^2x\left(Q(\partial u)^2
+\alpha' R^{(2)}\phi\right)}
is the action for a Feigin-Fuchs Coulomb gas, which is a one-dimensional
CFT with central charge given by
$c_1=1+6\alpha'(\partial\phi)^2$\ginsp. The imaginary charge of the
Feigin-Fuchs Coulomb gas describes the dilaton background growing
linearly in imaginary time and
$I_3$ is the Wess--Zumino--Witten\gepwit\ action on an $SU(2)$ group manifold
with central charge
\eqn\threecharge{c_3={3k\over k+2}\simeq 3-{6\over k}+{12\over k^2}+...}
where $k=Q/\alpha'$, called the ``level" of the WZW model, is an
integer. This can be seen from
the quantization condition on the Wess-Zumino term\gepwit
\eqn\iwzw{\eqalign{I_{WZ}&={i\over 4\pi\alpha'}\int_{\partial S_3}
d^2x\epsilon^{ab}\partial_ax^\mu\partial_bx^\nu B_{\mu\nu}\cr
&={i\over 12\pi\alpha'}\int_{S_3}d^3x\epsilon^{abc}
\partial_ax^\mu \partial_bx^\nu\partial_cx^\lambda H_{\mu\nu\lambda}\cr
&=2\pi i\left({Q\over\alpha'}\right).\cr}}
Thus $Q$ is not arbitrary, but is quantized in units of $\alpha'$.

We use this splitting to obtain exact expressions
for the fields by fixing the metric and antisymmetric tensor field
in their lowest order form and rescaling the dilaton order by order in
$\alpha'$. The resulting expression for the dilaton is
\eqn\alldilaton{e^{2\phi}={Q\over r^{\sqrt{{4\over 1+{2\alpha'\over Q}}}}}.}

We now turn to the heterotic multi-instanton solution of
\refs{\chsone,\chstwo}.
The tree-level supersymmetric vacuum equations for the heterotic string are
given by
\eqn\suei{\delta\psi_M=\left(\nabla_M-{\textstyle {1\over 4}}H_{MAB}\Gamma^{AB}
\right)\epsilon=0,}
\eqn\sueii{\delta\lambda=\left(\Gamma^A\partial_A\phi-{\textstyle{1\over 6}}
H_{AMC}\Gamma^{ABC}\right)\epsilon=0,}
\eqn\sueiii{\delta\chi=F_{AB}\Gamma^{AB}\epsilon=0,}
where $\psi_M,\ \lambda$ and $\chi$ are the gravitino, dilatino and gaugino
fields. The Bianchi identity is given by
\eqn\bianchi{dH=\alpha' \left(\tr R\wedge R-{\textstyle{1\over 30}}\Tr
        F\wedge F\right).}

The $(9+1)$-dimensional Majorana-Weyl fermions decompose down to
chiral spinors according to $SO(9,1)\supset SO(5,1) \otimes SO(4)$ for
the $M^{9,1}\to M^{5,1}\times M^4$ decomposition.
Let $\mu,\nu,\lambda,\sigma=1,2,3,4$ and $a,b=0,5,6,7,8,9$. Then the ansatz
\eqn\anstz{\eqalign{\met&=e^{2\phi}\delta_{\mu\nu},\cr g_{ab}&=\eta_{ab},\cr
H_{\mu\nu\lambda}&=\pm\epsilon_{\mu\nu\lambda\sigma}\partial^\sigma\phi\cr}}
with constant chiral spinors $\epsilon_\pm$ solves the supersymmetry
equations with zero background fermi fields provided the YM gauge
field satisfies the instanton (anti)self-duality condition
\eqn\yminst{F_{\mu\nu}=\pm {1\over 2}\epsilon_{\mu\nu}{}^{\lambda\sigma}
F_{\lambda\sigma}.}
An exact solution is obtained as follows. Define a generalized connection by
\eqn\genc{\Omega^{AB}_{\pm M}=\omega^{AB}_M\pm H^{AB}_M}
embedded in an SU(2) subgroup of the gauge group, and equate it
to the gauge connection $A_\mu$\dine\ so that $dH=0$ and the corresponding
curvature $R(\Omega_{\pm})$ cancels against the Yang-Mills field strength $F$.
As in the bosonic case, for $e^{-2\phi}\Box\ e^{2\phi}=0$ with the
above ansatz, the curvature of the generalized connection can be written in the
covariant form\rkinst
\eqn\gencurv{\eqalign{R(\Omega_\pm)_{\mu\nu}^{mn}
=&\delta_{n\nu}\nabla_m\nabla_\mu\phi
- \delta_{n\mu}\nabla_m\nabla_\nu\phi + \delta_{m\mu}\nabla_n\nabla_\nu\phi
- \delta_{m\nu}\nabla_n\nabla_\mu\phi \cr
&\pm \epsilon_{\mu mn\alpha}\nabla_\alpha\nabla_\nu\phi
\mp \epsilon_{\nu mn\alpha}\nabla_\alpha\nabla_\mu\phi ,\cr}}
from which it easily follows that
\eqn\gcinst{R(\Omega_\pm)^{mn}_{\mu\nu}=
\mp\half\epsilon_{\mu\nu}^{\ \ \ \lambda\sigma}
R(\Omega_{\pm})_{\lambda\sigma}^{mn}.}
Thus we have a solution with the ansatz \anstz\ such that
\eqn\exsol{F_{\mu\nu}^{mn}=R(\Omega_{\pm})_{\mu\nu}^{mn},}
where both $F$ and $R$ are (anti)self-dual.
This solution becomes exact since $A_\mu=\Omega_{\pm\mu}$
implies that all the higher order corrections
vanish\refs{\dine,\brone,\brtwo,\chsone,\chstwo,\rkdg}.
The self-dual solution for the gauge connection is then given by the 't Hooft
ansatz
\eqn\hfanstz{A_\mu=i \overline{\Sigma}_{\mu\nu}\partial_\nu \ln f.}
For a multi-instanton solution $f$ is again given by
\eqn\finst{f=e^{-2\phi_0}e^{2\phi}
=1+\sum_{i=1}^N{\rho_i^2\over |\vec x - \vec a_i|^2},}
where $\rho_i^2$ is the instanton scale size and $\vec a_i$ the location in
four-space of the $i$th instanton. An interesting feauture of the heterotic
solution is that it combines a YM instanton structure in the gauge
sector with an axionic instanton structure in the gravity sector.
In addition, the heterotic solution has finite action.

Note that the single instanton solution in the heterotic case
carries through to higher order without correction to the dilaton.
This seems to contradict the bosonic solution by suggesting that
the expansion for the Weyl anomaly coefficient $\wbd$ terminates at one loop.
This contradiction is resolved by noting that for a
supersymmetric ansatz the bosonic contribution to the
central charge is given by\swzw
\eqn\scharge{c_3={3k'\over k'+2}~,}
where $k'=k-2$. This reduces to
\eqn\termcharge{\eqalign{c_3&=3-{6\over k}\cr &=3-{6\alpha'\over Q},\cr}}
which indeed terminates at one loop order. The exactness of the splitting
then requires that $c_1$ not get any corrections from
$(\partial\Phi)^2$ so that $c_1+c_3=4$ is exact for the tree-level value
of the dilaton\refs{\chsone,\chstwo,\rkdg}.

\newsec{Three-Dimensional Monopole Solutions}

In this section we review the solutions with three-dimensional spherical
symmetry, and which have monopole-like structure. We begin with a simple
modification of the 't Hooft ansatz\refs{\rkmono,\rkmonex}\ which leads to a
multimonopole solution in field theory, not in the BPS limit\refs{\bogo,\prso}
and in itself far less interesting than the BPS solution. We then note that a
tree-level bosonic multi-soliton solution with monopole-like structure can be
written down\rkthes. Finally, we combine the gauge solution with the bosonic
solution to obtain an exact heterotic multimonopole
solution\refs{\rkmono,\rkmonex}.

We now return to the 't Hooft ansatz and the four-dimensional Euclidean action
\eqn\ymact{S=-{1\over 2g^2}\int d^4x {\rm Tr} G_{\mu\nu}G^{\mu\nu},
\qquad\qquad \mu,\nu =1,2,3,4}
with gauge group $SU(2)$.
We obtain a mutlimonopole solution by modifying the 't Hooft ansatz
\eqn\hfanstz{A_\mu=i \overline{\Sigma}_{\mu\nu}\partial_\nu \ln f}
as follows. We single out a direction in the transverse four-space (say $x_4$)
and assume all fields are independent of this coordinate. Then the solution
for $f$ satisfying $f^{-1} \Box\ f=0$ can be written as
\eqn\fmono{f=1+\sum_{i=1}^N{m_i\over |\vec x - \vec a_i|},}
where $m_i$ is the charge and $\vec a_i$ the location in
the three-space $(123)$ of the $i$th monopole. If we make the identification
$\Phi\equiv A_4$ (we loosely refer to this field as a Higgs field in
this paper, even though there is no apparent symmetry breaking mechanism),
then the Lagrangian density for the above ansatz can be rewritten as
\eqn\lgdn{\eqalign{G_{\mu\nu}^a G_{\mu\nu}^a =&G_{ij}^a G_{ij}^a +
2G_{k4}^a G_{k4}^a \cr =&G_{ij}^a G_{ij}^a + 2D_k \Phi^a D_k \Phi^a , \cr}}
which has the same form as the Lagrangian density for YM + massless scalar
field in three dimensions.

We now go to $3+1$ dimensions with the Lagrangian density (signature $(-+++)$)
\eqn\ymhlag{{\cal L}=-{1\over 4}G_{\mu\nu}^a G^{\mu\nu a} -{1\over 2}
D_\mu \Phi^a D^\mu \Phi^a,}
and show that the above multimonopole ansatz is a static solution with
$A_0^a=0$
and all time derivatives vanish. The equations of motion in this limit are
given by
\eqn\ymheqs{\eqalign{D_i G^{jia}&=g\epsilon^{abc}(D^j\Phi^b)\Phi^c,\cr
D_i D^i \Phi^a&=0.\cr}}
It is then straightforward to verify that the above equations are solved by
the modified 't Hooft ansatz
\eqn\monsol{\eqalign{\Phi^a&=\mp{1\over g}\delta^{ai}\partial_i \omega,\cr
A_k^a&=\epsilon^{akj}\partial_j \omega,\cr}}
where $\omega\equiv \ln f$. This solution represents a multimonopole
configuration with sources at $\vec a_i=1,2...N$. A simple observation of
far field and near field behaviour shows that this solution does not arise
in the Prasad-Sommerfield\prso\ limit. In particular, the fields are singular
near the sources and vanish as $r\to\infty$. This solution can be thought of
as a multi-line source instanton solution, each monopole being interpreted
as an ``instanton string''\rossi.

The topological charge of each source is easily computed
($\hat\Phi^a\equiv {\Phi^a/|\Phi|}$) to be
\eqn\topo{Q=\int d^3x k_0={1\over 8\pi}\int d^3x\epsilon_{ijk}\epsilon^{abc}
\partial_i\hat\Phi^a\partial_j\hat\Phi^b\partial_k\hat\Phi^c=1.}
The magnetic charge of each source is then given by $m_i=Q/g=1/g$.
It is also straightforward to show that the Bogomoln'yi\bogo\ bound
\eqn\bobo{G_{ij}^a=\epsilon_{ijk}D_k\Phi^a}
is saturated by this solution. Finally, it is easy to show that
the magnetic field $B_i={1\over 2}\epsilon_{ijk}F^{jk}$ (where
$F_{\mu\nu}\equiv \hat\Phi^a G_{\mu\nu}^a-(1/g)\epsilon^{abc}\hat\Phi^a
D_\mu \hat\Phi^b D_\nu \hat\Phi^c$ is the gauge-invariant electromagnetic
field tensor defined by 't Hooft\thooft) has the the far field limit behaviour
of a multimonopole configuration:
\eqn\bfield{B(\vec x)\to \sum_{i=1}^N {m_i(\vec x - \vec a_i)\over
|\vec x - \vec a_i|^3},\qquad {\rm as}\quad r\to \infty.}
As usual, the existence
of this static multimonopole solution owes to the cancellation of the
gauge and Higgs forces of exchange--the ``zero-force'' condition.

We have presented all the monopole properties of this solution.
Unfortunately, this solution as it stands has divergent
action near each source, and this singularity cannot be simply removed
by a unitary gauge transformation. This can be seen for a single source
by noting that as $r\to 0$, $A_k\to {1\over 2}\left(U^{-1}\partial_k U\right)$,
where $U$ is a unitary $2\times 2$ matrix. The expression in parentheses
represents a pure gauge, and there is no way to get around the $1/2$ factor in
attempting to ``gauge away'' the singularity\raj. The field theory
solution is therefore not very interesting physically. As we shall later in
this section, however, we can obtain an analogous finite action solution in
heterotic string theory. As in the previous section, we first consider a
monopole-like solution in bosonic string theory.

If we again single out a direction (say $x_4$) in the transverse space $(1234)$
of the bosonic string and assume all fields are independent of $x_4$, then
the tree-level bosonic multi-soliton solution to the string equations of
motion with the ansatz \sansatz\ is given by\rkthes\
\eqn\bomons{\eqalign{e^{2\phi}&=C+\sum_{i=1}^N
{m_i\over |\vec x -\vec a_i|},\cr
g_{\mu\nu}&=e^{2\phi}\delta_{\mu\nu},\qquad\qquad \mu,\nu=1,2,3,4,\cr
g_{ab}&=\eta_{ab},\qquad\qquad a,b=0,5,6...25,\cr
H_{\alpha\beta\gamma}&=\pm\epsilon_{\alpha\beta\gamma}{}^\mu\partial_\mu
\phi,\qquad \alpha,\beta,\gamma,\mu=1,2,3,4,\cr}}
where $\vec x=(x_1,x_2,x_3)$ is a three-dimensional coordinate
vector in the $(123)$ subspace of the transverse space. $m_i$ represents the
charge and $\vec a_i$ the location in the three-space of the $i$th source.

By singling out a direction $x_4$ and projecting out all
the field dependence on it, we destroy the $SO(4)$ invariance
in the transverse space possessed by the instanton solution\rkinst. However,
\bomons\ is an equally valid solution to the string equations as the
multi-instanton solution, since in both cases the dilaton field satisfies the
Poisson equation $e^{-2\phi}\Box\ e^{2\phi}=0$. The projection is necessary
to obtain the three-dimensional symmetry of a magnetic monopole.

Although the above bosonic multi-soliton solution \bomons\ lacks the gauge
and Higgs fields normally attributed to a magnetic monopole in field theory,
one can think of the dual field in the transverse four-space
$\hstar \equiv {1\over 6}\epsilon_{\alpha\beta\gamma\mu}
H^{\alpha\beta\gamma}$ as the magnetic field strength of a multimonopole
configuration in the space $(123)$ (note that $H^\ast_4=0$).

Unlike the four-dimensional solutions, the three-dimensional
solutions do not easily lend themselves to a CFT description, and it is
therefore difficult to go beyond $O(\alpha')$ in obtaining stringy
corrections to the tree-level fields. In \rkinst, the $O(\alpha')$ correction
was worked out for the special case of a single source with $C=0$. As in the
four-dimensional case, the metric and antisymmetric
tensor are unchanged to $O(\alpha')$, but the dilaton is corrected:
\eqn\bomona{e^{2\phi}={m\over r}\left(1-{\alpha' \over 8mr}\right).}
Unlike the four-dimensional solution, however,
the dilaton correction is not a simple rescaling of
the power of $r$ to order $\alpha'$. This fact is intimately connected
with the difficulty in formulating a CFT description of the three-dimensional
solution.

We now combine the above solutions to construct an exact multimonopole solution
of heterotic string theory. The derivation of this solution closely parallels
that of the multi-instanton solution reviewed in section 2, but
in this case, the solution possesses three-dimensional (rather than
four-dimensional) spherical symmetry near each source. Again the reduction is
effected by singling out a direction in the transverse space. An exact solution
is now given by
\eqn\anstz{\eqalign{\met&=e^{2\phi}\delta_{\mu\nu},\qquad g_{ab}=\eta_{ab},\cr
H_{\mu\nu\lambda}&=\pm\epsilon_{\mu\nu\lambda\sigma}\partial^\sigma\phi,\cr
e^{2\phi}&=e^{2\phi_0}f,\cr
A_\mu&=i \overline{\Sigma}_{\mu\nu}\partial_\nu \ln f,\cr}}
where in this case
\eqn\fdmono{f=1+\sum_{i=1}^N{m_i\over |\vec x - \vec a_i|},}
where $m_i$ is the charge and $\vec a_i$ the location in
the three-space $(123)$ of the $i$th monopole. If we again identify the
Higgs field as $\Phi\equiv A_4$, then the gauge and Higgs fields may be simply
written in terms of the dilaton as
\eqn\stmono{\eqalign{\Phi^a&=-{2\over g}\delta^{ia}\partial_i\phi,\cr
A_k^a&=-{2\over g}\epsilon^{akj}\partial_j\phi\cr}}
for the self-dual solution. For the anti-self-dual solution, the Higgs
field simply changes sign. Here $g$ is the YM coupling constant. Note
that $\phi_0$ drops out in \stmono.

The above solution (with the gravitational fields obtained
directly from \anstz\ and \fdmono) represents an exact multimonopole
solution of heterotic string theory and has the same structure in
the four-dimensional transverse space as the above multimonopole solution
of the YM + scalar field action. If we identify
the $(123)$ subspace of the transverse space as the space part of the
four-dimensional spacetime (with some toroidal compactification, similar
to that used in \jim) and take the timelike direction as the usual $X^0$,
then the monopole properties of the field theory solution carry
over directly into the string solution.

The string action contains a term $-\alpha' F^2$ which also diverges
as in the field theory solution. However, this divergence
is precisely cancelled by the term $\alpha' R^2(\Omega_\pm)$ in the
$O(\alpha')$
action. This result follows from the exactness condition
$A_\mu=\Omega_{\pm\mu}$
which leads to $dH=0$ and the vanishing of all higher order corrections
in $\alpha'$. Another way of seeing this is to consider the higher order
corrections to the bosonic action shown in \refs{\brone,\brtwo}. All such
terms contain the tensor $T_{MNPQ}$, a generalized curvature incorporating
both $R(\Omega_\pm)$ and $F$. The ansatz is constructed precisely so that this
tensor vanishes identically\refs{\rkinst,\rkdg}. The action thus reduces to
its finite lowest order form and can be calculated directly for a multi-source
solution from the expressions for the massless fields in the gravity sector.

The divergences in the gravitational sector in heterotic string theory thus
serve to cancel the divergences stemming from the field theory solution. This
solution thus provides an interesting example of how this type of cancellation
can occur in string theory, and supports the promise of string theory as a
finite theory of quantum gravity. Another point of interest is that the string
solution represents a supersymmetric multimonopole solution coupled to gravity,
whose zero-force condition in the gravity sector (cancellation of
the attractive gravitational force and repulsive antisymmetric field force)
arises as a direct result of the zero-force condition in the gauge sector
(cancellation of gauge and Higgs forces of exchange) once the gauge connection
and generalized connection are identified.

\newsec{Conclusion}

We classified some of the recently obtained higher-membrane solitonic solutions
of string theory according to the symmetry the solitons possess in the space
transverse to the membrane. We considered in this paper two such classes: those
with four-dimensional spehrical symmetry, and which possess instanton
structure,
and those with three-dimensional symmetry, which represent magnetic
monopole-like solutions in string theory.

We outlined in section 2 the 't Hooft ansatz for the Yang-Mills instanton,
and then turned to the bosonic tree-level axionic instanton solution of
\reyone,
and its exact extension for the case of a single instanton wormhole
solution\rkinst. A combination of the gauge instanton and axionic instanton
solutions led to an exact multi-instanton solution in heterotic string
theory\refs{\chsone,\chstwo\rkdg}.

In section 3 we considered some of the monopole-like solutions. In this
case, a combination of the modified 't Hooft ansatz\refs{\rkmono,\rkmonex}
and a bosonic three-dimensional solution\rkthes\ led to an exact heterotic
multimonopole solution\refs{\rkmono,\rkmonex}.
Unlike the instanton solutions, the monopole solutions
do not seem to be easily describable in terms of conformal field theories,
an unfortunate state of affairs from the point of view of string theory.
An interesting aspect of this solution, however, is that the YM divergences
of the modified 't Hooft ansatz solution are precisely cancelled in
the string theory solution by similar divergences in the gravity sector,
resulting in a finite action solution. This finding is
significant in that it represents an example of how string theory
incorporates gravity in such a way as to cancel infinities inherent in
gauge theories, thus supporting its promise as a finite theory of quantum
gravity.

Another class of solutions, which we did not consider here, are the
eight dimensional instanton\refs{\gksone\gkstwo{--}\gksthree}\ solutions
of string theory\refs{\dabhar\dghrr{--}\dfluthree}. In this case, however,
the exact extension is most natural in the context of a dual theory of
fundamental fivebranes, which has not yet been constructed.

%\bigbreak\bigskip\bigskip\centerline{{\bf Acknowledgements}}\nobreak

\vfil\eject
\listrefs
\bye